\documentclass{sigchi}

\RequirePackage{snapshot}

\CopyrightYear{2018}
\setcopyright{acmlicensed}
\conferenceinfo{AutomotiveUI'18,}{September 23-25, 2018, Toronto, Canada}
\acmPrice{\$15.00}

\toappear{
\urlstyle{leobig}
\normalsize
* Corresponding author: Lex Fridman, fridman@mit.edu\\
** Website: \url{https://hcai.mit.edu/minimalism}
\urlstyle{leo}
}

\usepackage{balance}       
\usepackage{graphics}      
\usepackage[T1]{fontenc}   
\usepackage{txfonts}
\usepackage{mathptmx}
\usepackage[pdflang={en-US},pdftex]{hyperref}
\usepackage{color}
\usepackage{booktabs}
\usepackage{textcomp}
\usepackage{placeins}
\usepackage{subcaption}
\usepackage{float}

\usepackage{microtype}        
\usepackage{ccicons}          

\usepackage{todonotes}

\def\plaintitle{Designing Toward Minimalism in Vehicle HMI}

\def\emptyauthor{}
\def\plainkeywords{Authors' choice; of terms; separated; by
  semicolons; include commas, within terms only; required.}

\makeatletter
\def\url@leostyle{%
  \@ifundefined{selectfont}{
    \def\UrlFont{\sf}
  }{
    \def\UrlFont{\small\bf\ttfamily}
  }}
\makeatother
\def\pprw{8.5in}
\def\pprh{11in}

\definecolor{linkColor}{RGB}{6,125,233}
\hypersetup{%
  pdftitle={\plaintitle},
  pdfauthor={\emptyauthor},
  pdfkeywords={\plainkeywords},
  pdfdisplaydoctitle=true, 
  bookmarksnumbered,
  pdfstartview={FitH},
  colorlinks,
  citecolor=black,
  filecolor=black,
  linkcolor=black,
  urlcolor=linkColor,
  breaklinks=true,
  hypertexnames=false
}

\begin{document}

\title{\plaintitle}

\renewcommand{\thefootnote}{\fnsymbol{footnote}}
\newcommand{\authorspace}{\hspace{0.2in}}
\author{
Julia Kindelsberger \authorspace
Lex Fridman\footnote[1]{Fun fun asdf;lj}  \authorspace
Michael Glazer \authorspace
Bryan Reimer\vspace{0.05in}\\
\affaddr{Massachusetts Institute of Technology}\\
\vspace{0.1in}
}

\maketitle

\begin{abstract}
  We propose that safe, beautiful, fulfilling vehicle HMI design must start from a rigorous consideration of minimalist design. Modern vehicles are changing from mechanical machines to mobile computing devices, similar to the change from landline phones to smartphones. We propose the approach of ``designing toward minimalism'', where we ask ``why?''  rather than ``why not?'' in choosing what information to display to the driver. We demonstrate this approach on an HMI case study of displaying vehicle speed. We first show that vehicle speed is what 87.6\% of people ask for. We then show, through an online study with 1,038 subjects and 22,950 videos, that humans can estimate ego-vehicle speed very well, especially at lower speeds. Thus, despite believing that we need this information, we may not. In this way, we demonstrate a systematic approach of questioning the fundamental assumptions of what information is essential for vehicle HMI.
\end{abstract}




\section{Introduction}

\vspace{0.1in}
\makebox[\columnwidth][c]{%
  \begin{minipage}{0.8\columnwidth}
    \centering
    \textit{``Everything should be made as simple as possible, but not simpler.'' - Albert Einstein}
  \end{minipage}}

A software-centric approach to automotive engineering is changing the function, design, and role of the modern vehicle
\cite{mossinger2010software}. The revolutionary aspect of this approach is revealed when it's coupled with cellular
connectivity that allows for instantaneous over-the-air updates to all major software-based functionalities of the
vehicle's perception, control, and HMI systems. Furthermore, with two-way cellular communication, continuous data
collection enables individual vehicle customization based on driver-specific behavioral information gathered in that
vehicle. A fully software-controlled HMI puts the power and freedom to continuously innovate in the hands of automotive
engineers and designers. The current trend in HMI design is toward complexity, toward adding information not removing it
\cite{feld2011automotive,khan2016cross, business, GM}. In contrast to this trend, we propose a minimalist approach to
vehicle HMI design: start from scratch and only add the absolute minimum that creates a safe and enjoyable driving
experience.


\begin{figure}[H]
  \centering
  \includegraphics[width=0.98\columnwidth]{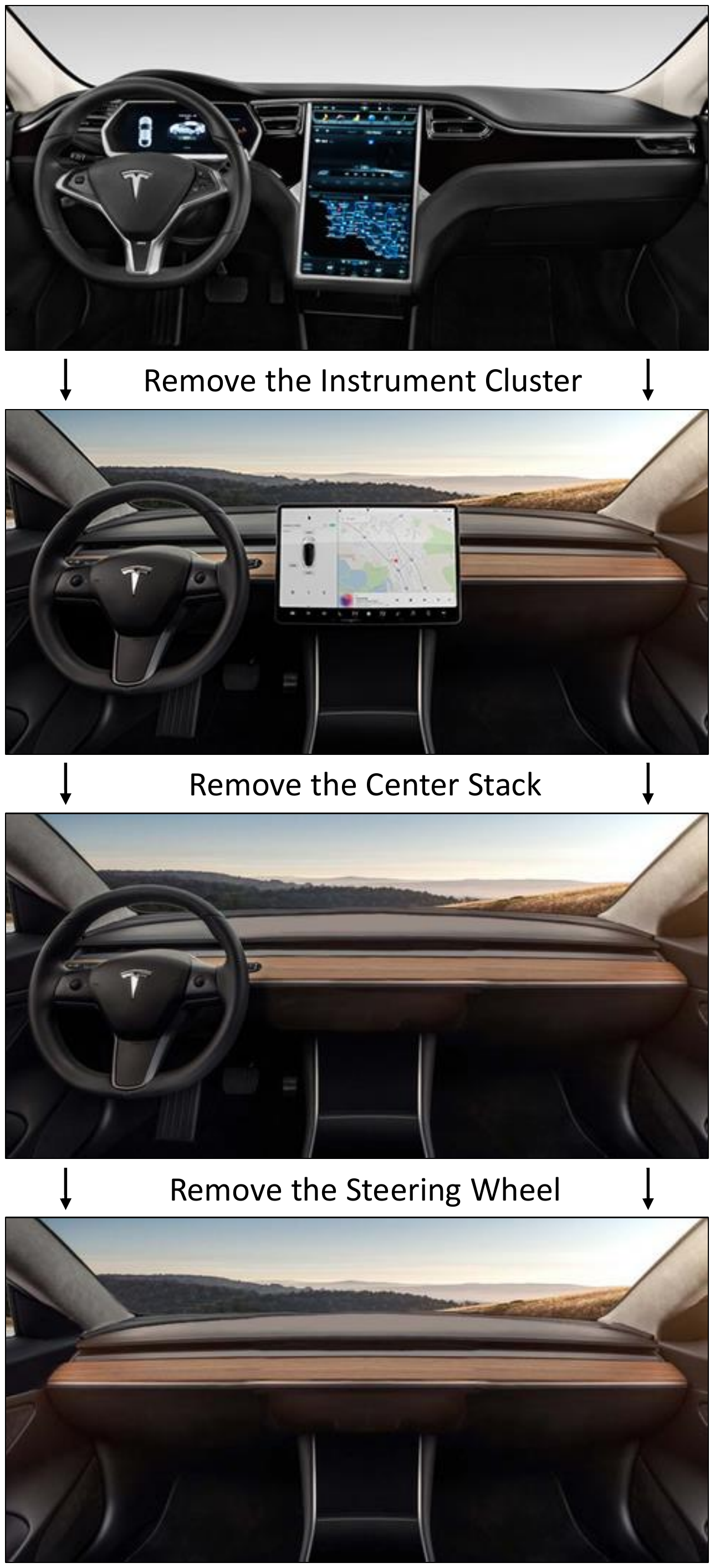}
  \caption{Visualization of ``designing toward minimalism'' in human-vehicle communication, using the design decision to
    remove the instrument cluster in the Tesla Model 3 as an illustrative initial step. This approach is characterized
    by starting from scratch and asking ``why?'' rather than ``why not?'' in choosing what information to add.}
  \label{fig:minimalism}
\end{figure}

Fig. ~\ref{fig:minimalism} is an illustration of designing toward minimalism, motivated by Tesla's decision not to include an instrument cluster in Tesla Model 3. We propose that removing the instrument cluster may be a step in the right direction, toward a more spatially-focused communication of information. The open question underlying the minimalist philosophy is: what else can be removed? From a pragmatic engineering perspective, the best way to systematically answer that question is by removing everything and only adding back the pieces of information that are provably essential. This paper is both a presentation of a design philosophy and a case study applying that philosophy to the question of whether we need to show vehicle speed at all. We show that vehicle speed is the most asked-for piece of information and we also show that there is evidence that we don't need it. Specifically, the main findings of this work in terms of the application of the minimalist philosophy to a specific case study of estimating speed are as follows:

\begin{enumerate}
\item 88\% of people find vehicle speed to be the most useful information while driving.
\item 80\% of speed estimations are within 10mph of actual speed. The error increases with greater speed.
\item 95\% of speed estimations in stop-and-go traffic are within 5mph of actual speed. This motivates context-dependent
  display of information.
\end{enumerate}

This paper presents the results of a questionnaire (N=170) about what information people look at the instrument cluster for and their opinions on removing it from the car. Since the speedometer is the HMI element that people claim to need the most while driving, in applying the minimalist design philosophy, this work presents two speed estimation experiments toward understanding what happens if we don't show the speed at all. Firstly, we present a speed experiment that shows participants 17 videos of vehicles driving in the first person perspective (N=960). The mean speed estimation error of this experiment is 0.19mph and 58\% of all estimations were below 5mph. To investigate whether people improve over time, a second experiment (N=78) showed participants 85 videos (5 of each speed). The mean estimation error of that experiment is 0.33mph and over 60\% of all estimations were within 5mph. Participants were better in estimating lower speeds. Taking into account that a driver taking their eyes off the road (e.g., to look at the instrument cluster) is a factor for many accidents, removing the instrument cluster removes sources for driver distraction. Ultimately, removing all visual information from the car, including the center stack, encourages drivers to do the only thing they should be doing while driving: focus on the road ahead.

\begin{figure}
\begin{center}
 \includegraphics[width=\columnwidth]{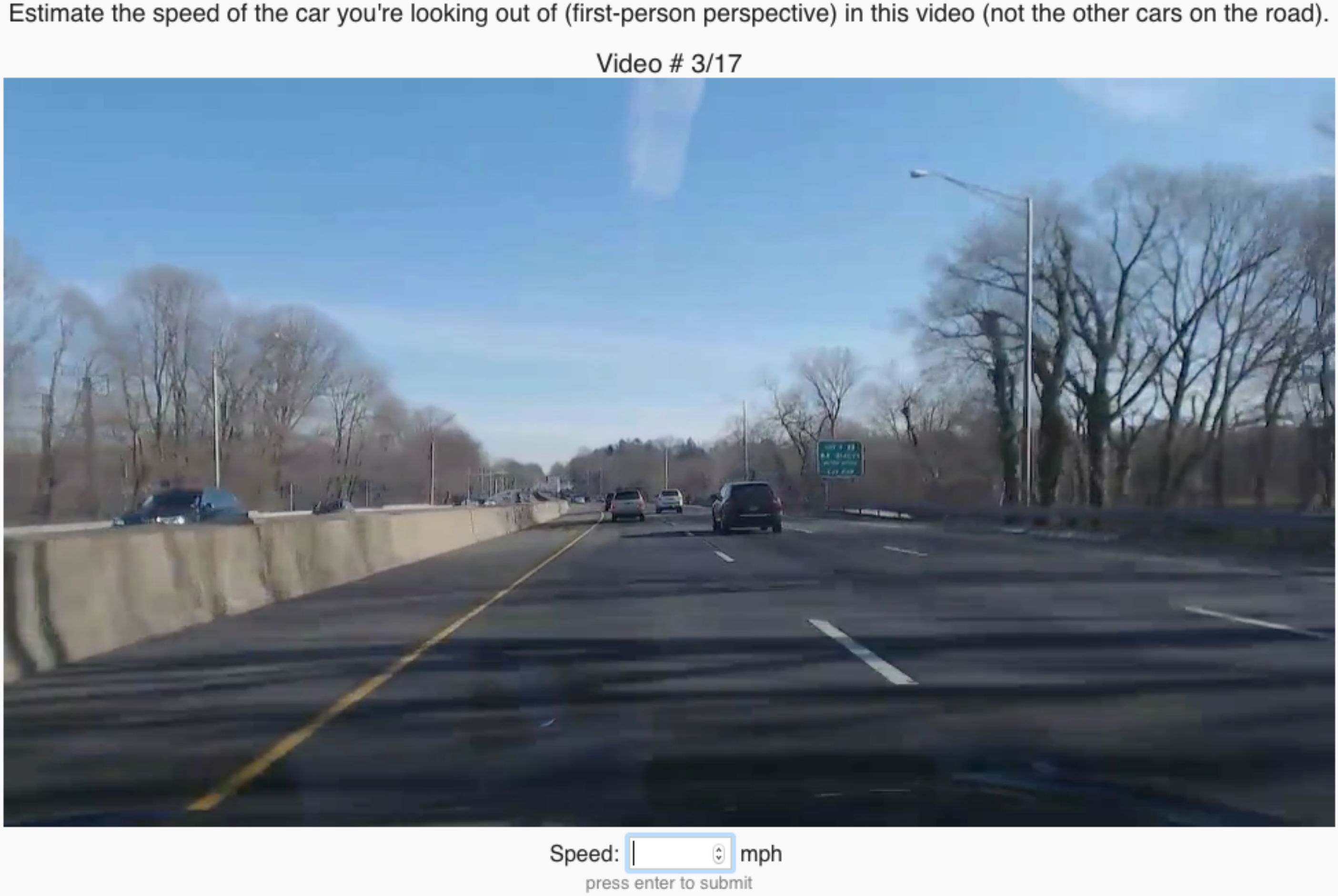}
 \caption{This image shows the user interface for the speed estimation experiments. It displays a video of the road from the driver's perspective and asks participants to estimate the speed of the vehicle.}
 \label{fig:ss}
 \end{center}
\end{figure}

\section{Related Work}\label{sec:related-work}


This section outlines three related areas: Previous work in the fields of instrument cluster design, the human's ability to estimate vehicle speed and minimalism design in various fields. Previous research ~\cite{fridman2016driver} found that the instrument cluster is the most common reason for drivers to take their eyes off the road. Therefore, removing the possibility to be distracted by information displayed on the instrument cluster could decrease the number of times when drivers take their eyes off the road. Studies have looked at different interface options for the instrument cluster design and present a design space that describes where to position input and output devices in a car ~\cite{kern2009design}. Furthermore, studies have focused on the usability of the instrument cluster for certain target groups, e.g. older drivers  ~\cite{kim2011usability}. While FMVSS 101 guides controls and displays for vehicles \cite{auto}, the requirements for instrument clusters are shifting with fully autonomous cars and thus demand a review of how, where and when information may be called upon to facilitate safe and effective communication of information between the vehicle, environment and driver. Finally, there are proposals for instrument cluster design approaches for the changing experience of autonomous driving ~\cite{gowda2014dashboard, Byton, benderius2017best}. The minimalism design approach has been explored in various fields. The general approach in the minimalism design philosophy is to find the important features, rank them by importance and add what is necessary. Among others, this philosophy has been applied in art and design ~\cite{vaneenoo2011minimalism}, advertising ~\cite{margariti2017tauypology} and information visualization ~\cite{tufte2014visual}. The human ability to estimate vehicle speed has been studied with questionnaires, in driving simulators and in cars with covered instrument clusters. A summary of previous vehicle speed estimation experiments, with their number of participants, estimates, speeds and the overall estimation error is shown in Table ~\ref{tab:table1}. These experiments mainly focus on how accurate people are in estimating vehicle speed under different conditions and show its dependency on tasks, road type, gender, experience and other factors. In general, studies do not appear to investigate the replicability of reports across multiple trials or assess how perception my change over them with performance cuing (e.g. actual speed is revealed to them after responses). Overall, it appears that speed estimation is robust within a 5-10mph window of error.


\begin{table*}[t]
  \centering
  \caption{This table summarizes previous speed estimation experiments by describing the method, the experiment, the number of participants, the number of estimations, the number of distinct speeds and the estimation error.}
  \label{tab:table1}
  \begin{tabular}{ccccccc}
    \toprule
    Reference & Method &Experiment & \# Participants & \# Estimates  & Speeds (mph) & Error (mph) \\
    \midrule
    Evans \cite{evans1970speed} & Road & normal passenger & 18 & 44 & 10-60 & -1.64\\
    Evans \cite{evans1970speed} & Road & unable to see & 18 & 44 & 10-60 & 0.71\\
    Evans \cite{evans1970speed} & Road & diminished hearing & 18  & 44 & 10-60 & -4.7\\
    Evans \cite{evans1970speed} & Road & both & 18 & 44 & 10-60 & -5.68\\
    Stanisa et al. \cite{milos?evic1990speed} & Road & speed in curves, sign 50km/h & 135 & ~25 & 25 & -1.42\\
    Stanisa et al. \cite{milos?evic1990speed} & Road & speed in curves, sign 60km/h & 71 & ~25 & 25 &-2.8 \\
    Recarte et al. \cite{recarte1996perception} & Road & speed estimation & 60 & 4 & ~35-75 &-9.19\\
    Recarte et al. \cite{recarte1996perception} & Road & speed production & 30 & 4 & ~35-75 &-7.89 \\
    Haglund et al. \cite{haglund2000speed} & Road & measuring on highways & 533 & 1 & ~90 &-1.06 \\
    Hurwitz et al. \cite{hurwitz2005speed} & Simulator & speed at checkpoints & 8 & 20 & ~15-45 &-1.06 \\
    Hurwitz et al. \cite{hurwitz2005speed} & Road & speed at checkpoints & 8 & 20 & ~15-45 & 0.30 \\
    Svenson et al. \cite{svenson2012braking} & Questionnaire & braking stop sign & 61 & 8 & ~18-80 & -21.44 \\
    Svenson et al. \cite{svenson2012braking} & Questionnaire & braking child & 115 & 16 & ~15-45 & -17.05\\
    Svenson et al. \cite{svenson2012braking} & Questionnaire & braking line & 115 & 16 & ~15-45 & -22.06\\
    Svenson et al. \cite{svenson2012braking} & Questionnaire & speed at checkpoints & 89 & 8 & ~15-45 & -13.97 \\
    \bottomrule

  \end{tabular}
\end{table*}

\section{Case Study: Instrument Cluster Information and Speed Estimation}\label{sec:case-study}

Tesla started the production of its first full-electric sedan for the mass-market - the Tesla Model 3 - in July 2017 \cite{model3}, adding another model to their line-up, the SUV Tesla Model X and the luxury sedan Tesla Model S. One major change is the lack of a traditional instrument cluster or head-up display (HUD) to display information (such as the speed of the car, distance traveled, turning indicators, fuel gauge, maintenance indicators) in front of the driver. Instead, some of this information is displayed on a 15-inch touchscreen in the middle of the dash. The instrument cluster has been a standard fitting in (almost) all vehicles from 1910 onwards. This drastic design change performed by Tesla inevitably raises the following question: Is this a step towards removing all visual information from cars?

The instrument cluster in cars has evolved over time. Traditionally, the HMI elements that are needed for primary and secondary tasks are located behind or on the steering wheel and elements for tertiary tasks are located at the center stack \cite{tonnis2006survey}. The speedometer and warning lights were the first instruments put in front of the driver. Later, temperature gauges and fuel gauge have been added to the instrument cluster  \cite{knoll2017some}. Over time digital instruments have been added, such as displays that show route guidance, parking aid, gear indication, the time etc. Over the last decade, the information on the instrument cluster has been almost doubled \cite{gkouskos2014drivers}. Recently, digital instrument panels evolved that display information like speed, media controls and indicator lights, depending on the situation (e.g. driving vs. parking) \cite{instrumentpanel}. The density of information has changed but is still primarily presented centrally in front of the driver. Studies have shown that the risk for accidents increases when drivers take their eyes off the road \cite{klauer2006impact} and that the instrument cluster is a common reason to do so~\cite{fridman2016driver}. In order to understand what the instances are that people need the instrument cluster for while driving, this paper presents a study that asks people instrument cluster and center stack related questions and their opinions on Tesla's new approach. Participants responded that they need the instrument cluster the most for checking the speed and that they look at the speedometer in order to make sure that they are not speeding. 

In addition, they don't think that it is sufficient to have all instrument cluster related information in the middle of the dash, and do not like the idea of removing the instrument cluster. 

The next step is to research whether people actually need that information contained in the speedometer as much as reported, or if they are able to obtain the information without the help of a display, or are able to learn speed estimation sufficiently to perform the task. 


\begin{figure}
\begin{center}
 \includegraphics[width=\columnwidth]{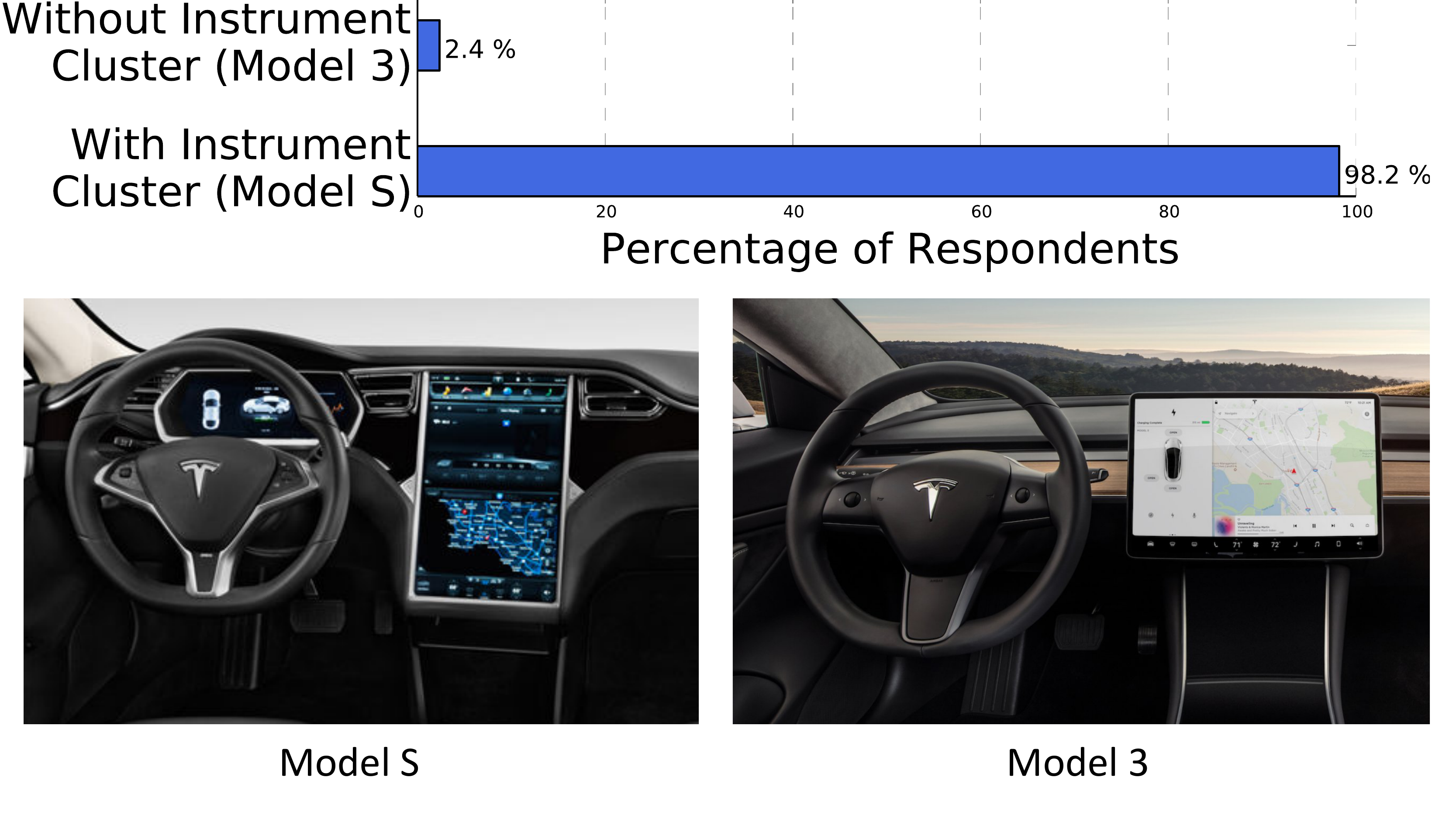}
 \caption{This chart shows the preferred layout of the space in front of the driver (Tesla Model S vs. Model 3) for providing the information needed while driving.}
 \label{fig:question_4}
 \end{center}
\end{figure}

\begin{figure}
\begin{center}
 \includegraphics[width=\columnwidth]{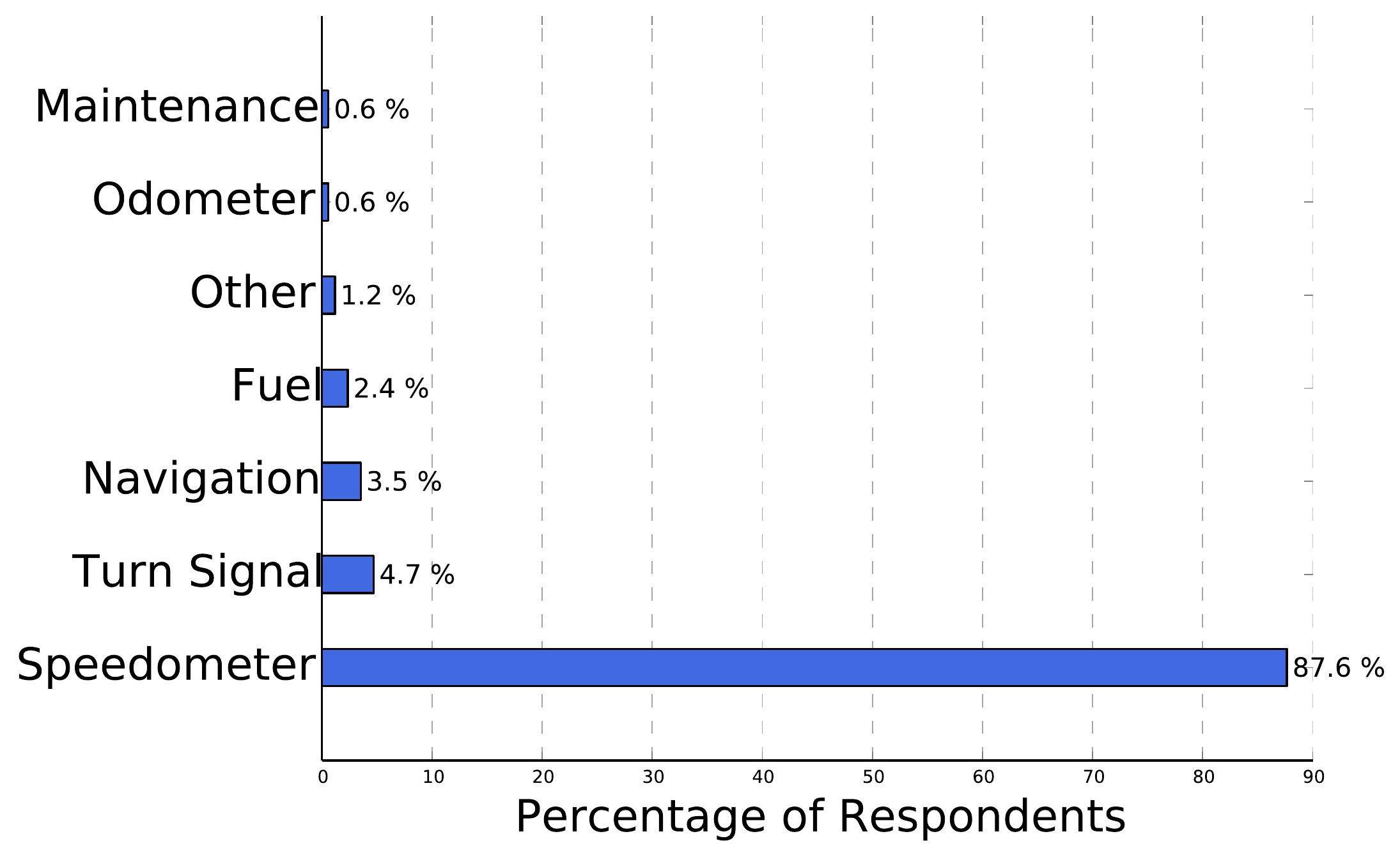}
 \caption{This image shows the user interface for the speed estimation experiments. It displays a video of the road from the driver's perspective and asks participants to estimate the speed of the vehicle.}
 \label{fig:question_2}
 \end{center}
\end{figure}

\begin{enumerate}
\item What do people think they need the instrument cluster for?
\item Do people need the instrument cluster for these tasks?
\end{enumerate}

\section{Instrument Cluster Questionnaire}
The main goal of this questionnaire was to: (1) develop a greater understanding of the type of information  people think they need in a vehicle instrument cluster; (2) assess what respondents think about the new approach of the Tesla Model 3 of removing the instrument cluster; and (3) consider to what extent they believe the instrument cluster or the center stack distracts them while driving.

To explore these concepts, we developed 10 questions that were field across 200 respondents. The questionnaire included one control question to make sure that participants were paying attention. This question showed the subjects six images, four of instrument clusters and two of center stacks and asked them to select all images of instrument clusters. Furthermore, we excluded participants that took less then 30 seconds to answer all questions. In total 170 participants passed both requirements. Moreover, participants were told that the questionnaire would involve answering questions related to the instrument cluster and the center stack of a car and those terms were described briefly. The questionnaire was put on Amazon Mechanical Turk, where participants had 5 minutes to answer all questions and got 0.5\$ after completing the survey. The full survey instrument is included in Appendix A.

The participants were asked two questions on what information they use while they drive: 1.) "What information do you use while you drive?" where they could select all that applied and 2.) "What information do you use the most while you drive?" where they could select one response. The provided response options were: "Speedometer (speed of the car)", "odometer (distance traveled)", "fuel gauge (how much gas/charge is left)", "turn signal indicators", "malfunction/maintenance indicators", "navigation" and "other". When they selected "other", they had to further specify. For the first question everyone, except one participant, included the speedometer in their responses. Furthermore, fuel gauge and turn signal indicators were included in 97\% and 93\% of answers, respectively. 66\% chose the malfunction/maintenance indicators, 49\% the odometer (distance traveled) and 41\% navigation. For the second question, 87\% of the participants chose the speed of the car as the information that they need the instrument cluster the most for (Fig.~\ref{fig:question_2}) followed by the turn signal indicators (5\%) and the navigation (4\%).
 
Participants were also shown two pictures, one of the space in front of the driver of the Tesla Model S and one of the Tesla Model 3, and asked which they would prefer for providing the information they selected in the previous questions while driving. 167 participants (98\%) prefer the Tesla Model S over the Tesla Model 3 (Fig.~\ref{fig:question_4}).

\begin{figure*}[th!]
  \begin{center}
  \includegraphics[width=\textwidth]{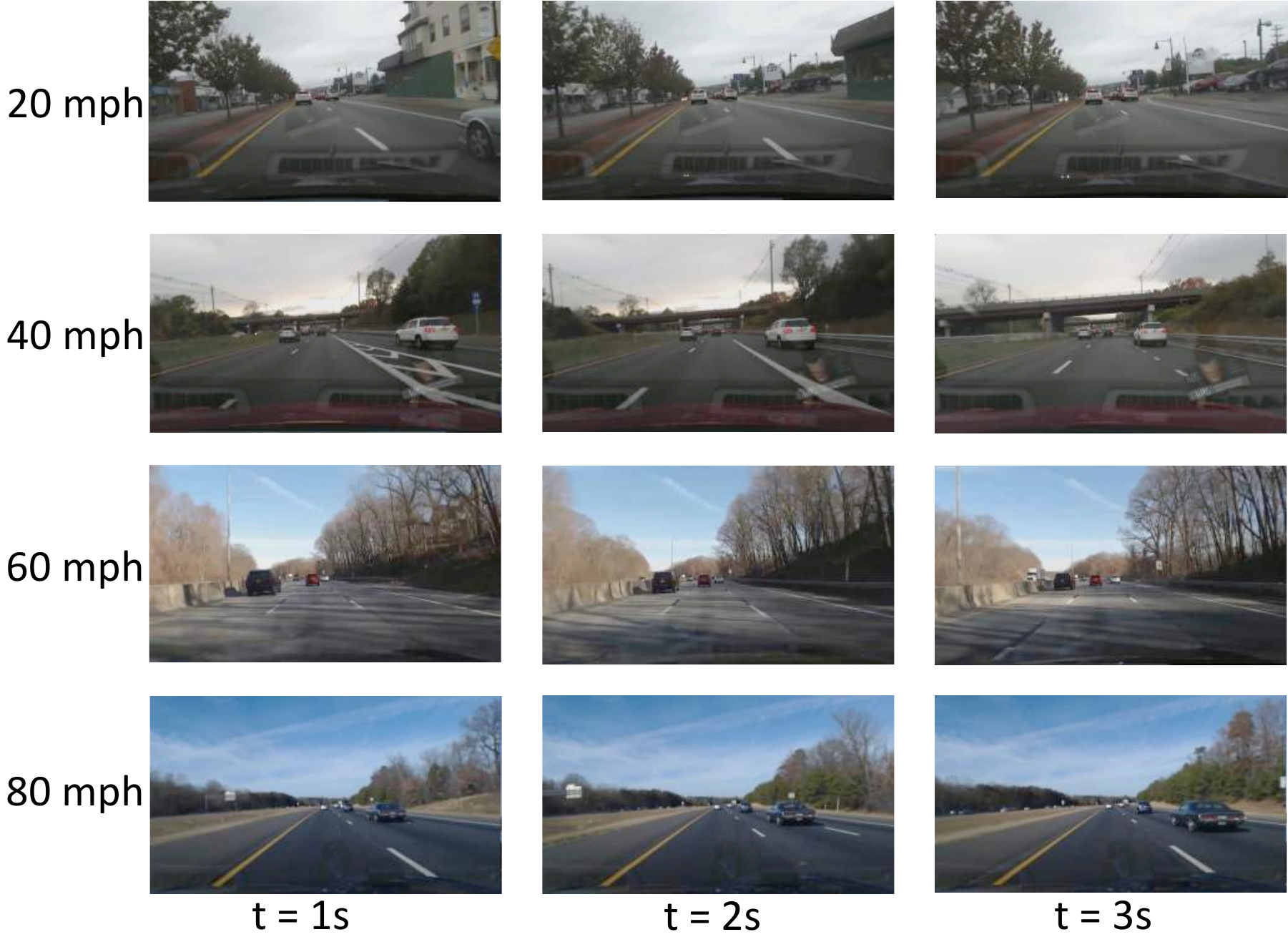}
  \caption{This visualization shows example videos of different speeds from the driver's perspective that were used for the speed estimation experiments.}
    \label{fig:video_image}
  \end{center}
\end{figure*}  

Furthermore, a large portion of the participants make sure that they are not speeding by looking at the instrument cluster (87\%). The provided response options were: "Look at the traffic", "look at the instrument cluster", "estimate the speed based on experience", "I don't care if I'm speeding or not" and "other". The next question, "Do you think it is helpful to have all instrument cluster related information in the center console?", was answered with "No" by 64\% of the participants. On average participants would spend 999\$ to upgrade the Tesla Model 3 with a traditionally placed (in front of the steering wheel) instrument cluster or a head-up display (not taking the three outliers into account) and 59\% would keep it as it is. 

Finally, participants generally don't think the instrument cluster distracts them while driving (91\%). Participants that answered this question with "Yes" had the option to further specify why. Five participants elaborated, four of them referred to lights on the instrument cluster that can be distracting ("If a warning indicator starts blinking", "Flashing lights, screen changes", "At night it could be too bright sometimes", "If unnecessary dash lights are on"). One participant mentioned alarm tones to be distracting ("Alarm going off"). Similarly, 77\% of the participants don't think that the center stack distracts them while driving. Some participants mentioned that they think looking to the side is more distracting than looking at the instrument cluster behind the wheel (e.g. "Don't like to look to the side. Much prefer the cluster right in front of the driving wheel").

The results suggest that people don't like the idea of removing the instrument cluster from the car and moving certain information to the middle of the dash of the car. Moreover, they responded that they mainly need the instrument cluster for checking the speed and to make sure that they are not speeding. However, they generally don't believe that the instrument cluster or the center stack distracts them while driving. One reason why participants oppose the idea of removing the instrument cluster might come from the fact that people tend to prefer what they are used to \cite{zajonc2001mere}. The question remains: to what extent do they actually need the provided information.

\section{Speed Survey}
Since the vehicle speed of the car is the information participants claimed to need the most while driving, we conducted a large-scale experiment to know how well participants are able to estimate vehicle speed and whether participants improve over time if the correct travel speed is revealed after each estimation. In contrast to existing speed estimation experiments, we went large-scale with the number as participants, range of vehicle speeds and the number of estimations. For this purpose, 850 videos (4 seconds long) that show a car on a road from a driver's perspective with different speeds (ranging from 0 mph, 5 mph, 10 mph ... 80 mph) were chosen from a large pool of videos that are available from several different cars over many different trips across the US. We generated 50 videos for each speed. Fig.~\ref{fig:video_image} shows frames of these videos for different speeds over time. A website was created that explained the experiment and showed participants videos for 4 seconds. These videos are randomly chosen from the pool of videos. After each video the participants were asked to estimate the speed of the vehicle they are looking out of (first person perspective) in mph and type the estimation into a textfield. After submitting their estimation, the actual speed is shown to the participants. Then, the next video automatically plays and this repeats until all videos are evaluated. The experiments were conducted on Amazon Mechanical Turk. For both experiments, we excluded participants that had the 0mph videos off by more than 3mph and were off by more than 40mph once.

\subsection{Speed Estimation Experiment 1}
\begin{figure}
\begin{center}
  \includegraphics[width=\columnwidth]{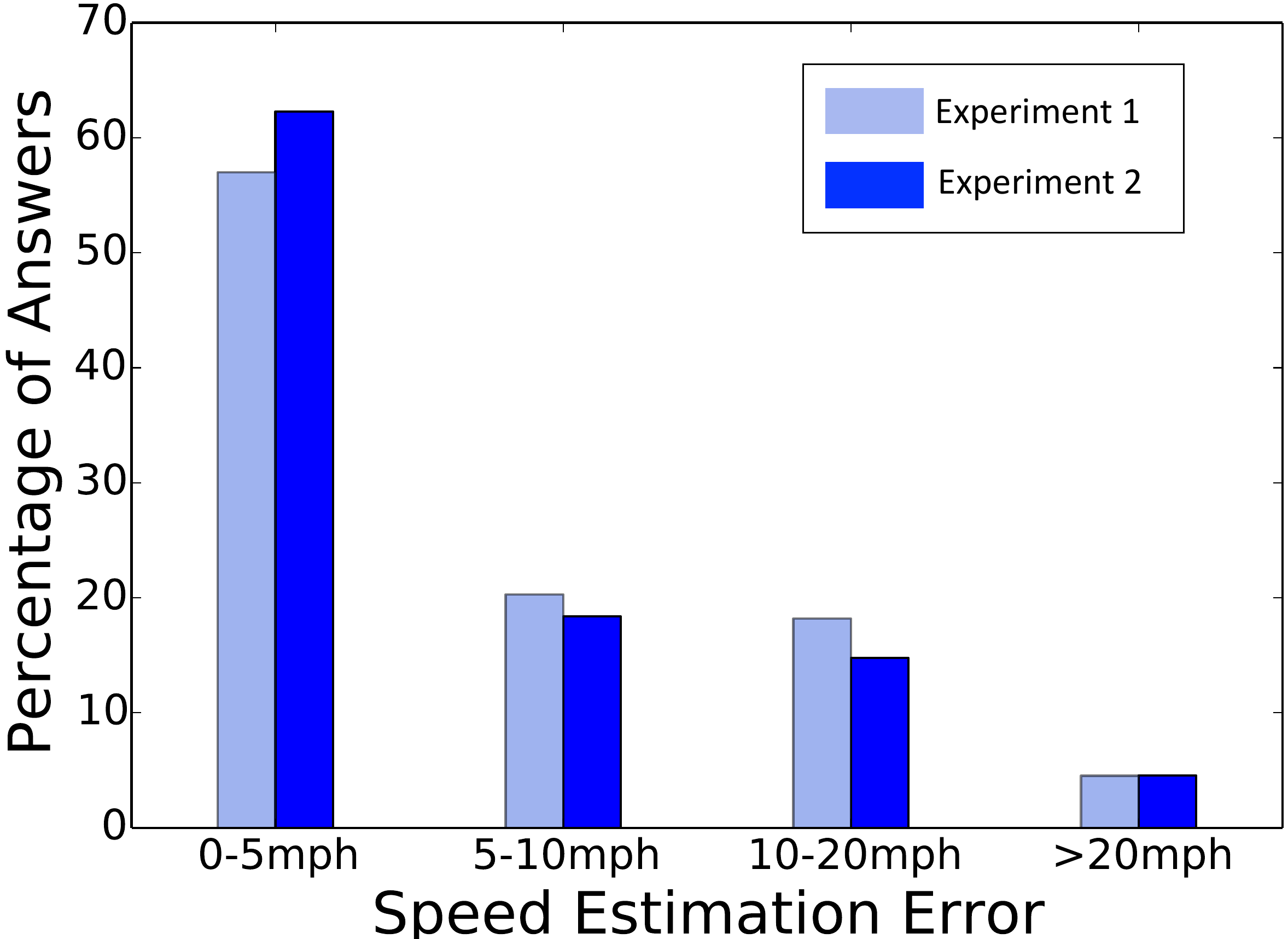}
  \caption{This chart shows the speed estimation error of all estimations for both speed estimation experiments.}
   \label{fig:estimation_count}
  \end{center}
\end{figure}

The main purpose of the first experiment was to study how accurately a large number of participants (N=960) are able to estimate the speed of moving vehicles over a large range of different speeds (0 - 80mph). In this experiment participants had 15 minutes to estimate the speed of 17 videos (one per each speed) and received 0.2\$. 1280 people participated. 25\% were eliminated based on the requirements described above, therefore the final sample includes 960 participants. 16,320 unique data points were created. 

The results show that 57\% of all estimations were within 5mph, 20.3\% between 5 - 10mph and 18.2\% between 10-20mph (Fig.~\ref{fig:estimation_count}). The absolute mean estimation error of all participants is 7.78mph and the relative mean error is 0.25mph. The estimation error for each participant is shown in Appendix B. Fig.~\ref{fig:speed3_heatmap_image}a and b show the relative and absolute speed estimation error for all shown speeds and show that participants are in general better in estimating lower speeds. The following estimations were within 5mph: 94.73\% of estimations <= 5mph, 77\% of estimations <= 20mph, but just 63.8\% of estimations <= 50mph. The absolute and relative mean estimation error for all participants in the order of videos is shown in Fig.~\ref{fig:speed3_heatmap_image}c and d. 22\% of all responses were correct responses (off by 0mph).

\begin{figure}
\begin{center}
  \includegraphics[width=\columnwidth]{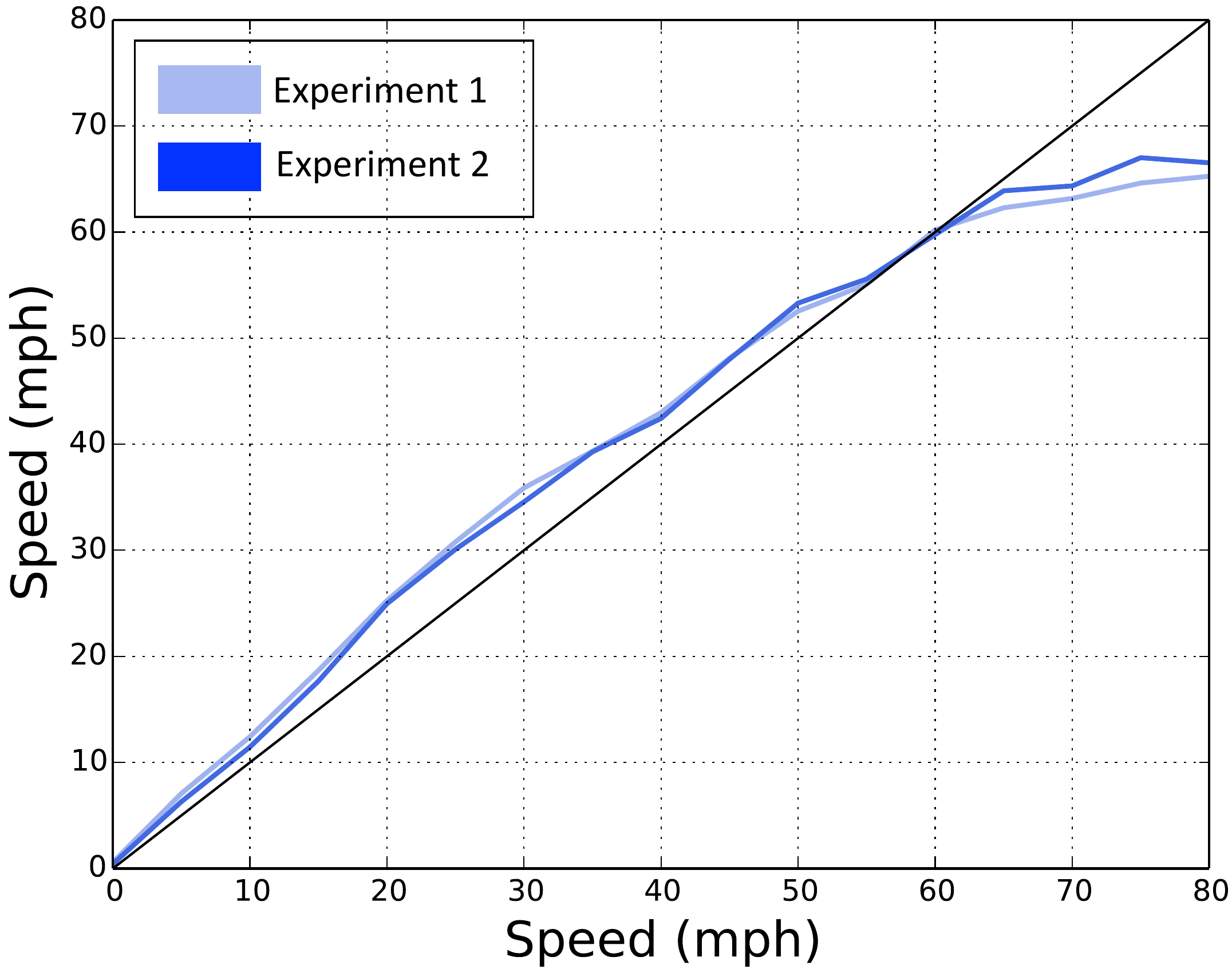}
  \caption{This chart shows the actual and estimated speed for both experiments. Participants overestimate lower speeds and underestimate higher speeds in both experiments.}
   \label{fig:act_est}
  \end{center}
\end{figure} 

  
  

This experiment shows participants were, in general, better in estimating lower speeds. The speed range from 20mph -  70mph was estimated similarly and speeds > 70mph were generally underestimated Fig.~\ref{fig:act_est}. The mean error is similar to previous studies (around 0mph). 

\begin{table}[th!]
  \centering
  \caption{This table shows the actual and average estimated vehicle speeds for both speed estimation experiments.}
  \label{tab:table2}
  \begin{tabular}{ccccccc}
    \toprule
    Speed (mph) & Experiment 1 (mph) & Experiment 2 (mph)\\
      \midrule
    0 & 0.59 &  0.42 \\
    5 & 7.07 & 6.3 \\
    10 & 12.42 & 11.45 \\
    15 & 18.67 & 17.66 \\
    20 & 25.26 & 24.96 \\
    25 & 30.79 & 30.09 \\
    30 & 35.88 & 34.56 \\
    35 & 39.34 & 39.28 \\
    40 & 42.98 & 42.43 \\
    45 & 48.1 & 48.02 \\
    50 & 52.54 & 53.3 \\
    55 & 55.07 & 55.57\\
    60 & 60.2 & 59.76\\
    65 & 62.29 & 63.89\\
    70 & 63.17 & 64.35\\
    75 & 64.62 & 67.01\\
    80 & 65.26 & 66.53\\
    \bottomrule
  \end{tabular}
\end{table}

\subsection{Speed Estimation Experiment 2}
To evaluate whether people improve over time, a longer experiment with 85 videos per speed was conducted. This experiment produced 6630 unique data points.

Participants had 30 minutes to estimate the speed of 85 videos (5 of each speed) and they received 2\$ for completing all videos. 90 people participated and 78 participants passed the two requirements.

The results show that 6\% of the participants were able to estimate vehicle speed with a mean absolute estimation error of < 5mph. The mean absolute estimation error and the mean error are slightly lower than in the first experiment: 7.31mph and 0.33 mph and the absolute error of 60\% of the participants is below the mean. Furthermore, a large portion, 62\%, of the absolute error of the estimations is under 5 mph (Fig.~\ref{fig:estimation_count}), 81\% under 10 mph and  91\% under 15mph.

\begin{figure*}[th!]
\begin{center}
  \includegraphics[width=13cm]{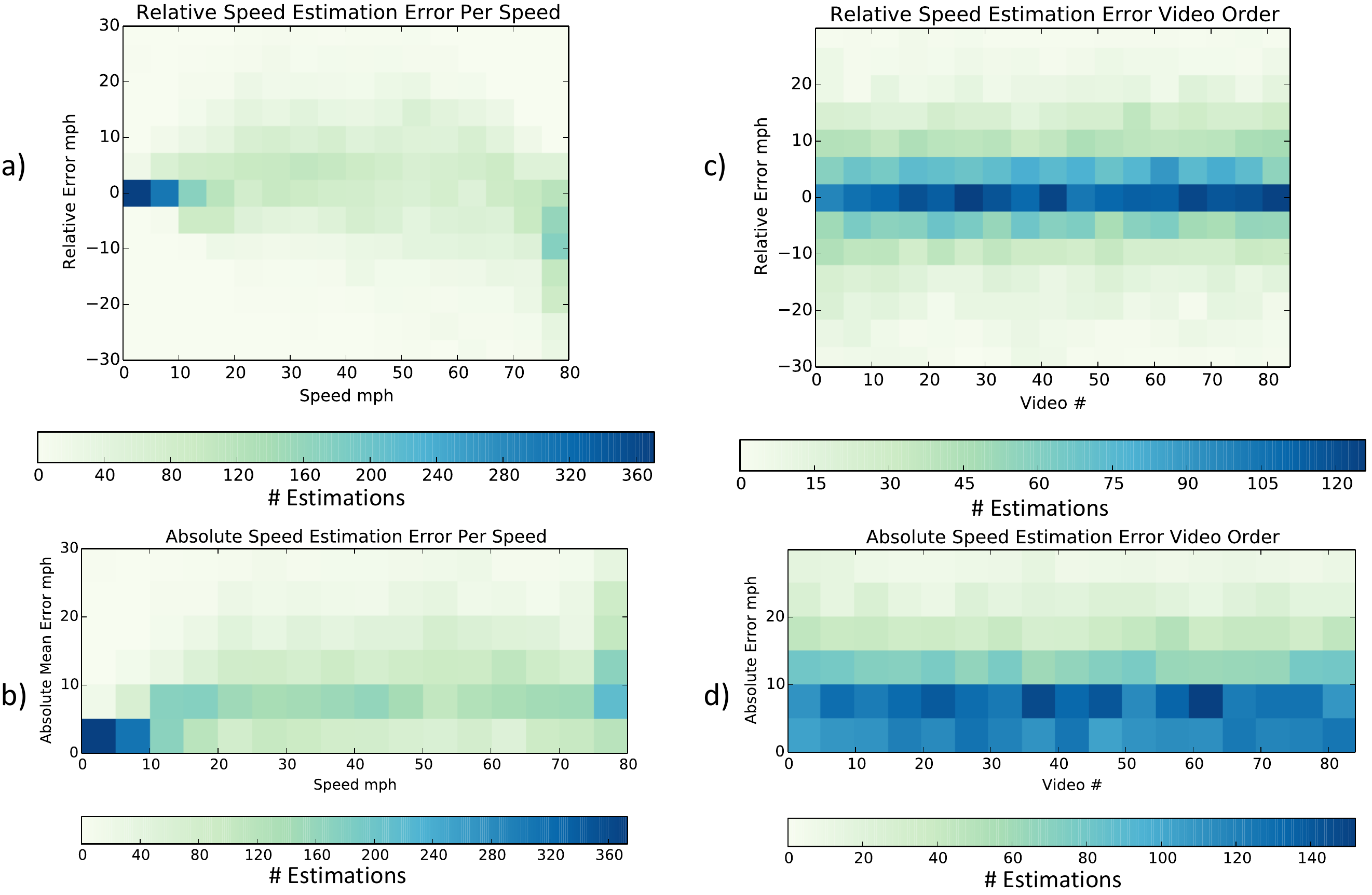}
  \caption{Experiment 2: Heat maps that show the mean relative speed estimation error per speed of all estimations (a), the mean absolute estimation error per speed (b), the mean relative estimation error in the order of the videos shown to the participants (c) and the absolute estimation error in the order of the videos shown to the participants (d).}
    \label{fig:speed2_heatmap_image}
  \end{center}
\end{figure*}

Similar to the first experiment, participants were generally better in estimating the vehicle speed at lower speeds. When looking at the different speeds, 96\% of 0 - 5 mph, 79\% of 5 - 10 mph and 30\% of 10 - 15 mph were estimated with an absolute estimation error below 5 mph. The absolute mean estimation error per speed is shown in Fig.~\ref{fig:speed2_heatmap_image}b. When looking at the relative speed error (Fig.~\ref{fig:speed2_heatmap_image}a), there is sightly more overestimation, except for very high speeds (75 - 80mph). The mean estimation error decreased slightly from 7.4 to 7.2 mph when looking at the first and second half of the experiment. The absolute and relative mean estimation error for all participants in the order of videos is shown in Fig.~\ref{fig:speed2_heatmap_image}c and ~\ref{fig:speed2_heatmap_image}d. The correct responses increased: almost a third (29\%) of all responses were correct responses (off by 0mph). When looking at the individual participants, some participants improved over time especially after the first couple of videos and some participants perform steady throughout.

\begin{figure*}[th!]
\begin{center}
  \includegraphics[width=13cm]{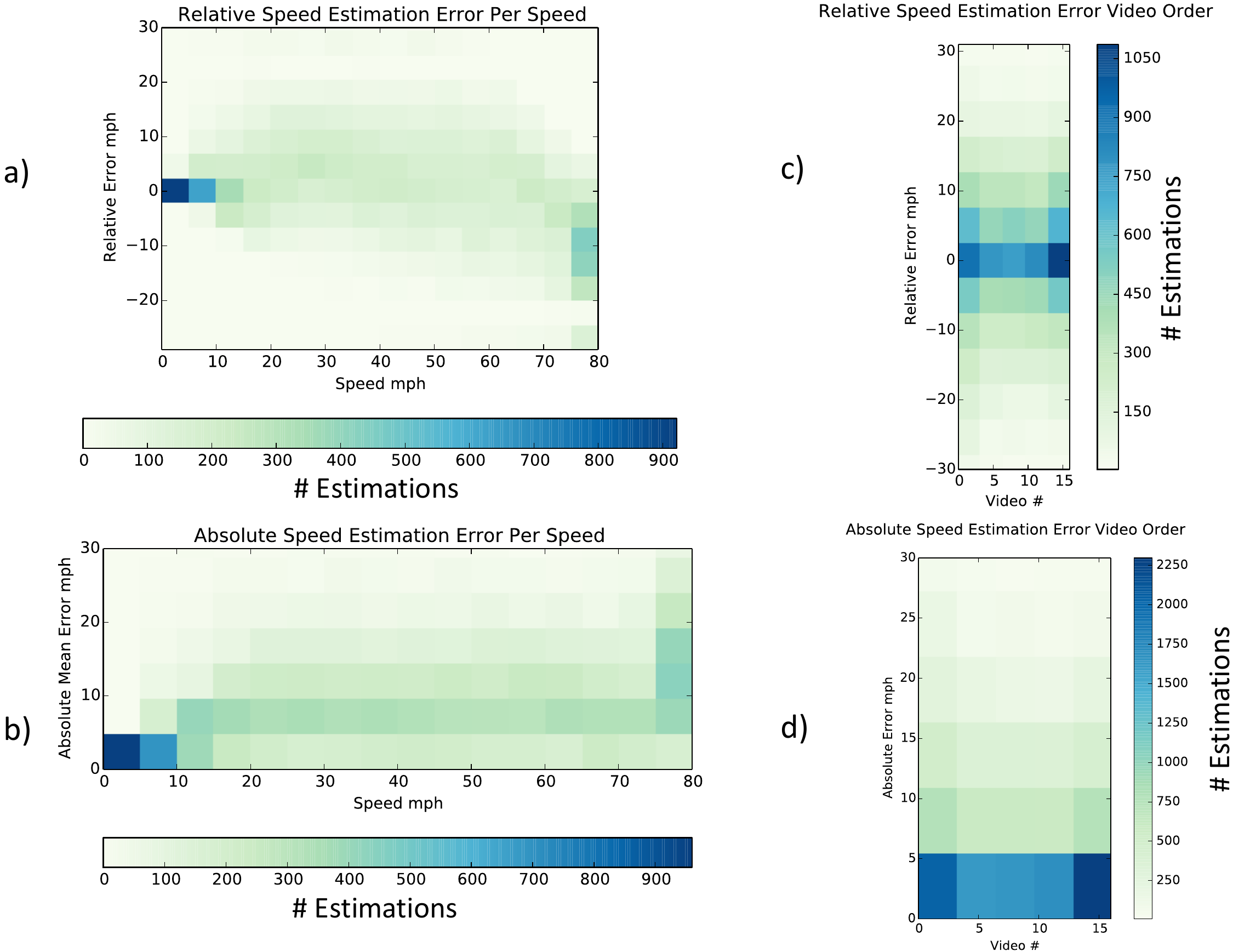}
  \caption{Experiment 1: Heat maps that show the mean relative speed estimation error per speed of all estimations (a), the mean absolute estimation error per speed (b), the mean relative estimation error in the order of the videos shown to the participants (c) and the absolute estimation error in the order of the videos shown to the participants (d).}
  \label{fig:speed3_heatmap_image}
  \end{center}
\end{figure*}

The errors in the second experiment are slightly lower than in the first one and the correct answers are slightly higher. This might indicate a learning ability of estimating the speed of a moving vehicle. The estimation errors of both experiments for both experiments all speeds are summarized in Fig. ~\ref{fig:act_est}.

\section{Proposed Design}\label{sec:proposed}
\begin{figure}
  \includegraphics[width=\columnwidth]{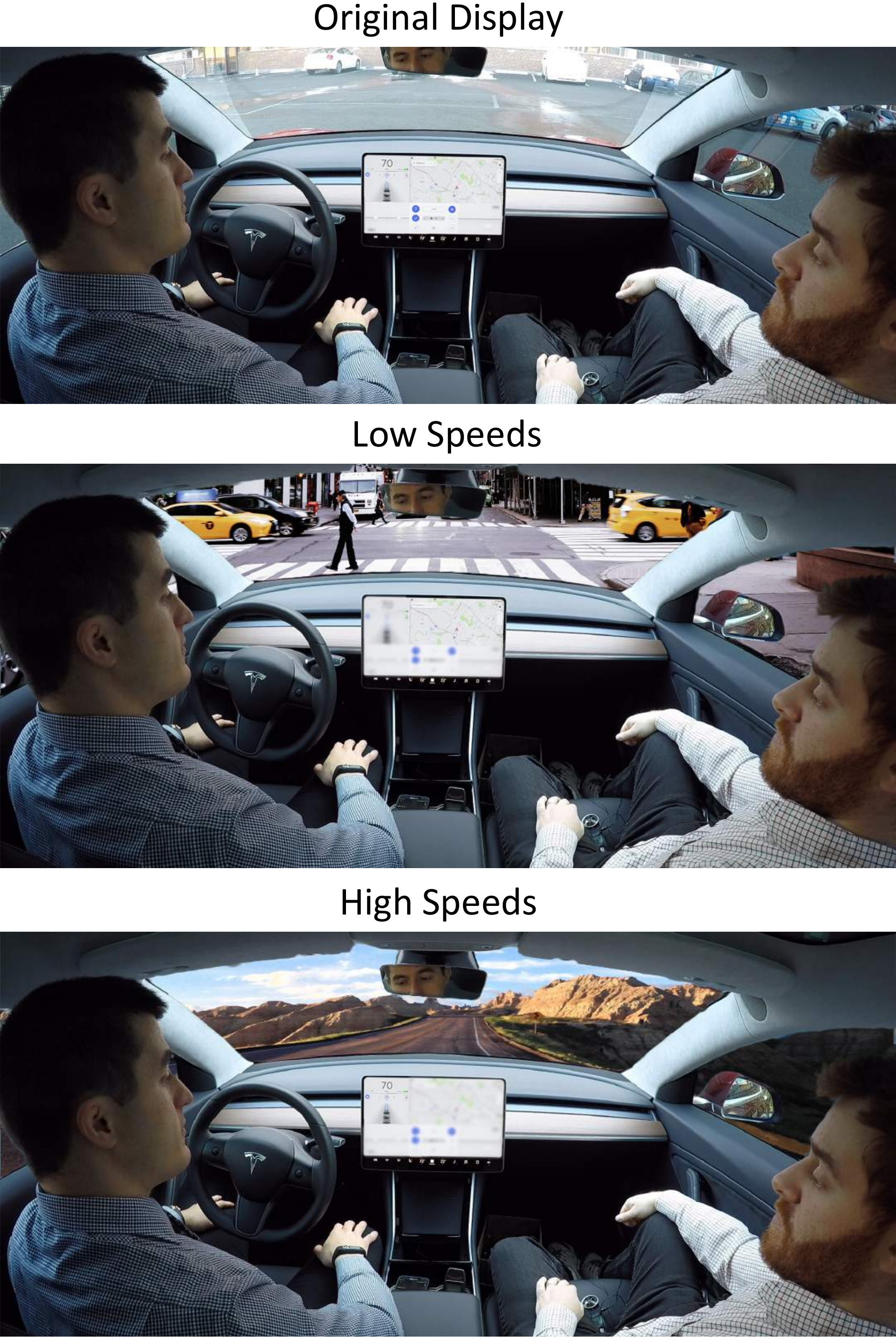}
  \caption{This image shows the application of the minimalism design philosophy. The first image shows the original display of a Tesla Model 3. The second image shows the display while driving in urban environments, where the speed is not displayed to the driver. The third image shows the display when driving with high speeds, where the speed is displayed to the driver.}
  \label{fig:image_new}
\end{figure}

We propose a minimalism design that shows as little information as possible to the driver. Our speed estimation case study shows that people are better in estimating lower speeds. Therefore, one application of our philosophy is to only show the speed above a certain speed, since the speed may be a source of distraction when driving in an urban environment. Fig. ~\ref{fig:image_new} illustrates this approach: The first image shows the original display in a Tesla Model 3. The second image shows the display when driving with low speeds, there the part of the display that shows the speed is blurred to avoid distraction. The third image shows the display for high speeds where the speed is displayed to the driver.

\section{Conclusion}\label{sec:conclusion}

This work presents and applies the philosophy of minimalist design to modern vehicle HMI. As part of the application of
this philosophy, we first showed that people think they need the instrument cluster to check the speedometer in order to
make sure that they are not speeding. We then conducted two experiments that revealed that a large percentage of drivers are able to accurately
estimate speed within 5mph and thus do not need the HMI to present this information to them. These results suggest drivers may not need this information presented in the center stack to successfully maintain an appropriate speed. In fact, the presence of this information may be a source of distraction when considered in the context of the self-reported need to check it frequently, though these results need to be validated within an on-road study. We propose a minimalism design that shows as little information as possible to the driver. In future work, we will seek to further apply the minimalist design philosophy toward the display of automation-related information to study driver response and behavior 'in the wild'.    

\balance{} 

\bibliographystyle{sigchi}
\bibliography{minimalism}

\clearpage

\section{Appendix A}\label{sec:appendix}
\begin{figure}[H]
\begin{center}
  \includegraphics[width=7cm]{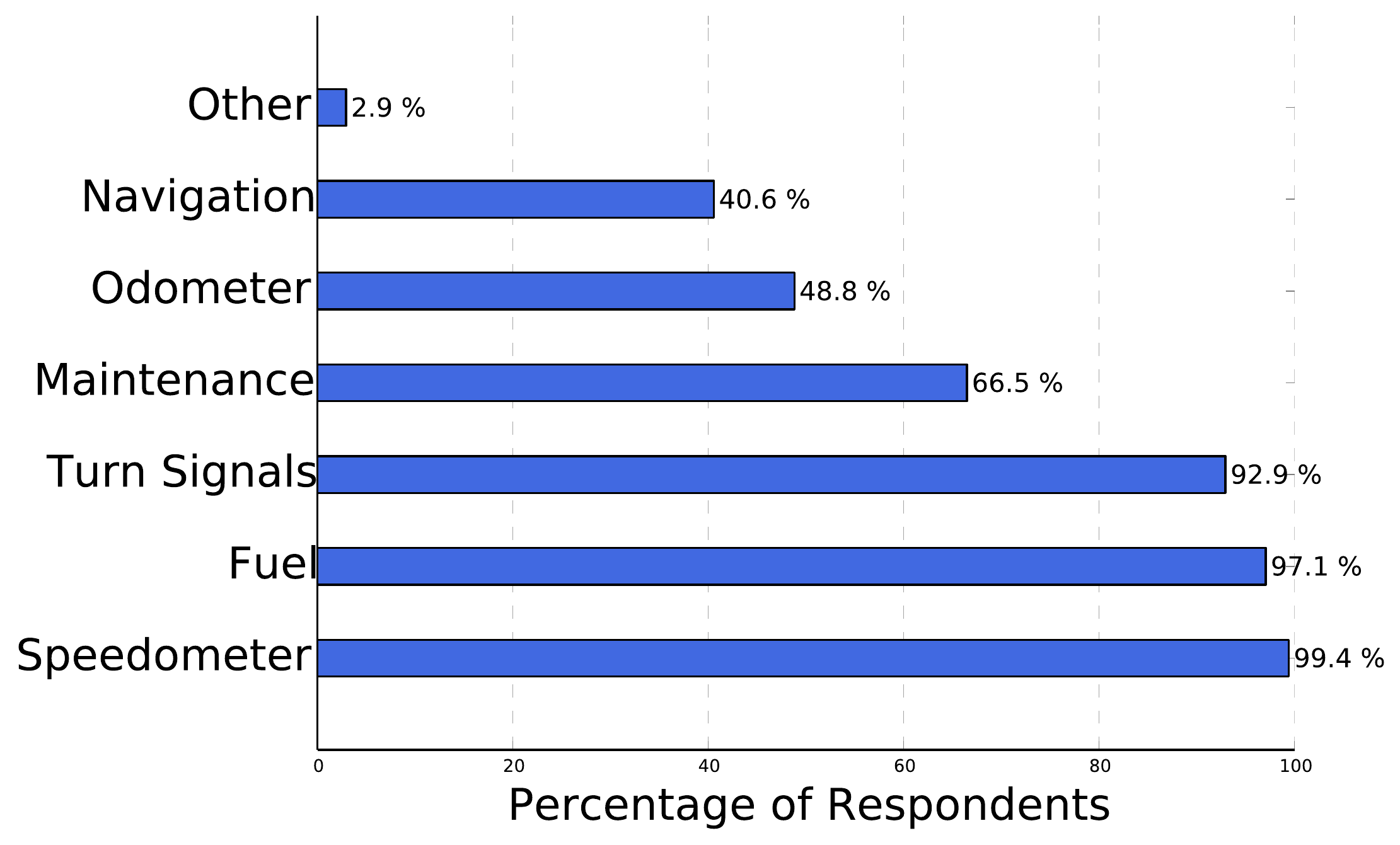}
  \caption{What information do you use while you drive?}
  \label{fig:1}
  \end{center}
\end{figure}

\begin{figure}[H]
\begin{center}
  \includegraphics[width=7cm]{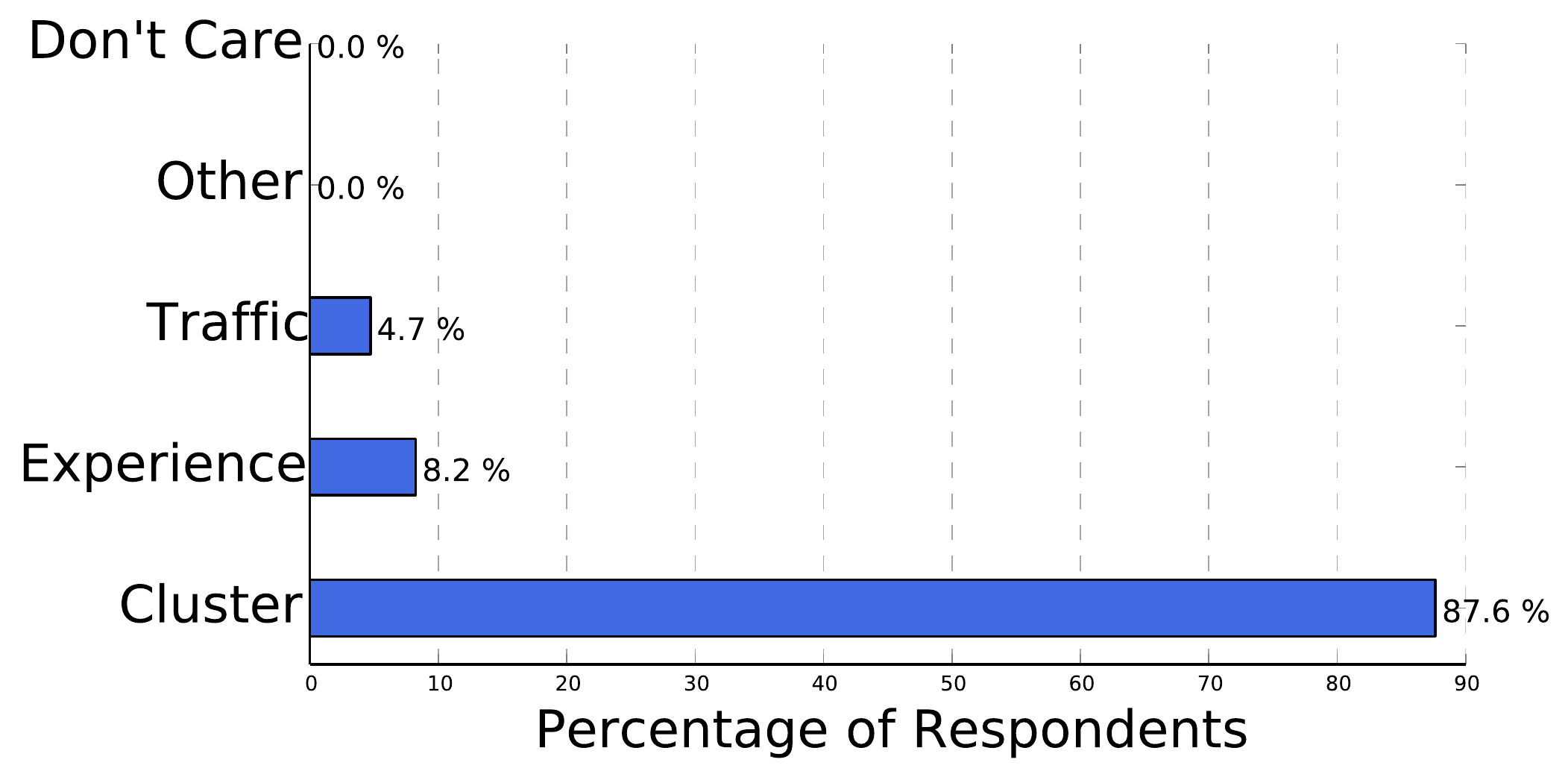}
  \caption{How do you ensure you are not speeding while driving?}
   \label{fig:5}
  \end{center}
\end{figure}

\begin{figure}[H]
\begin{center}
  \includegraphics[width=7cm]{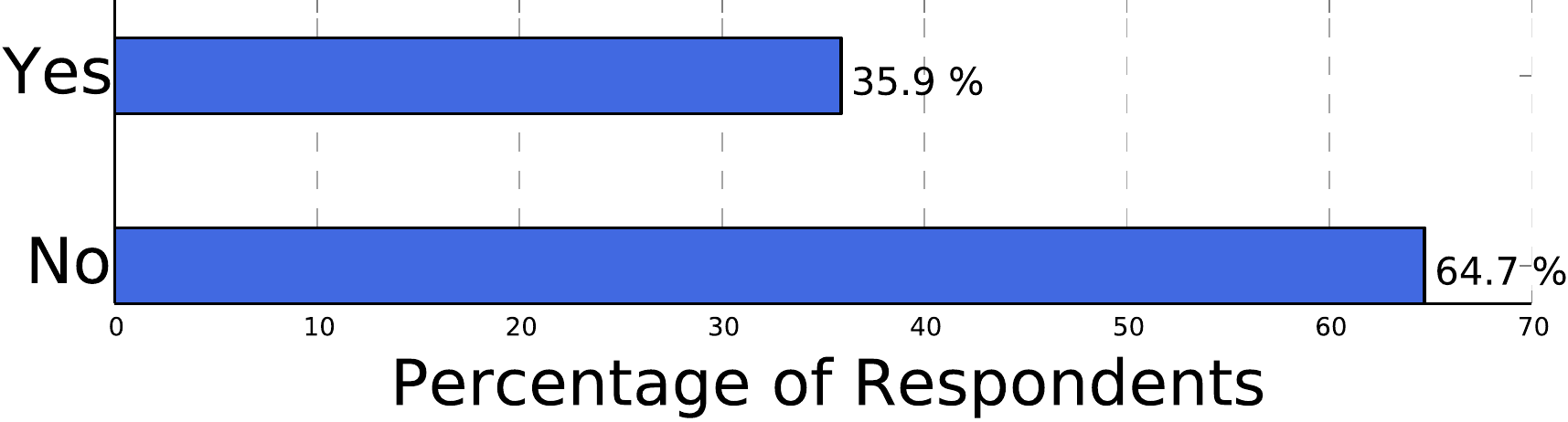}
  \caption{Do you think it is helpful to have all instrument cluster related information in the center console?}
   \label{fig:7}
  \end{center}
\end{figure}

\begin{figure}[H]
\begin{center}
  \includegraphics[width=7cm]{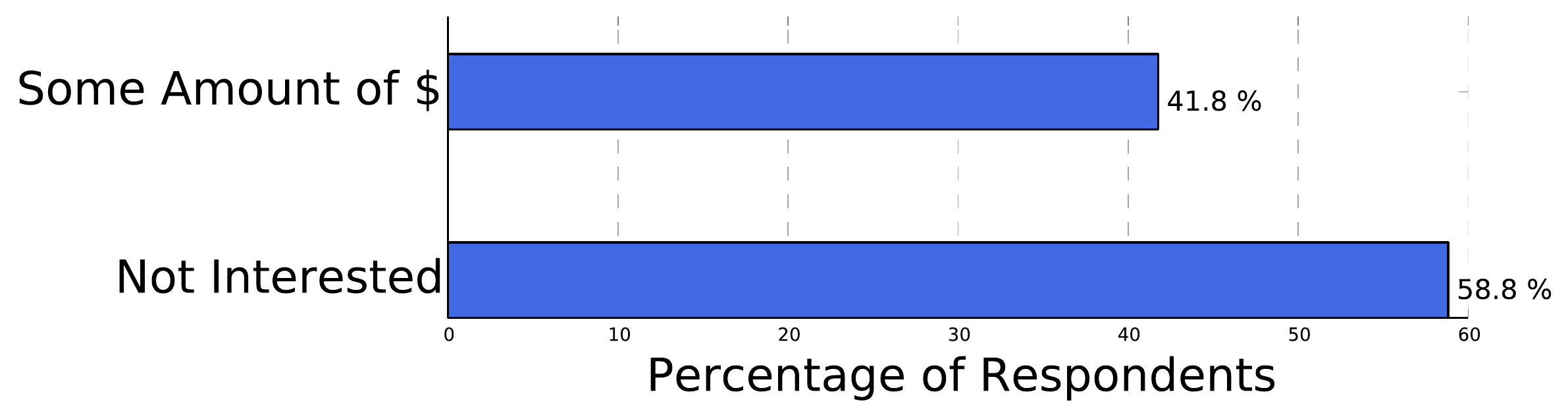}
  \caption{How much would you pay to upgrade your new Tesla Model 3 with a traditionally placed (in front of the steering wheel) instrument cluster of head-up display?}
  \label{fig:8}
  \end{center}
\end{figure}

\begin{figure}[H]
\begin{center}
  \includegraphics[width=7cm]{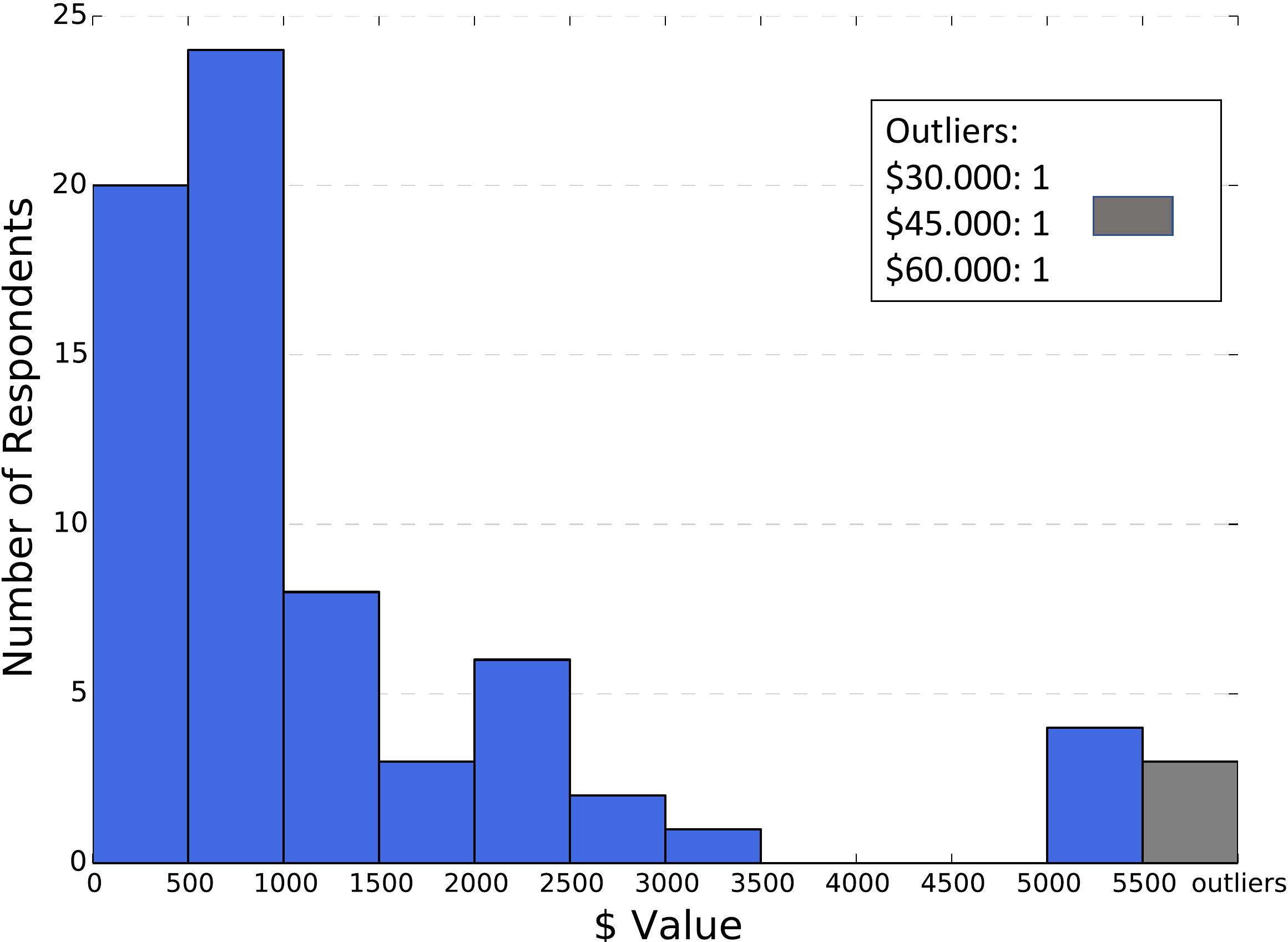}
  \caption{This histogram how much people would pay to upgrade the Tesla Model 3 with a traditionally placed instrument cluster or HUD.}
   \label{fig:8h}
  \end{center}
\end{figure}

\begin{figure}[H]
\begin{center}
  \includegraphics[width=7cm]{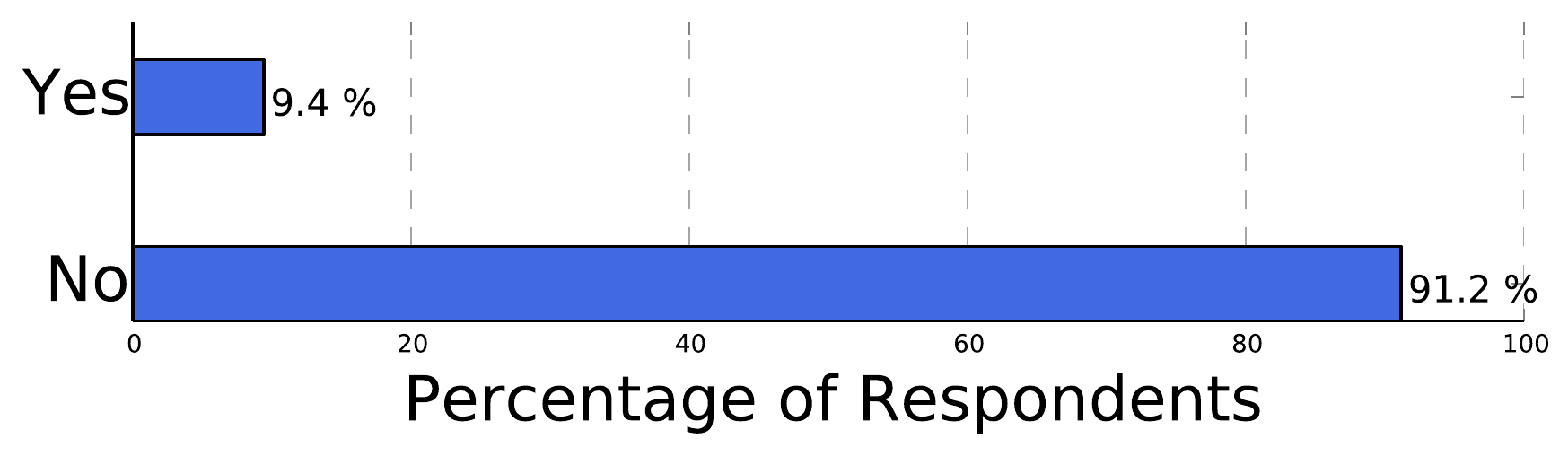}
  \caption{Are there any instances in which the instrument cluster distracts you?}
   \label{fig:9}
  \end{center}
\end{figure}

\begin{figure}[H]
\begin{center}
  \includegraphics[width=7cm]{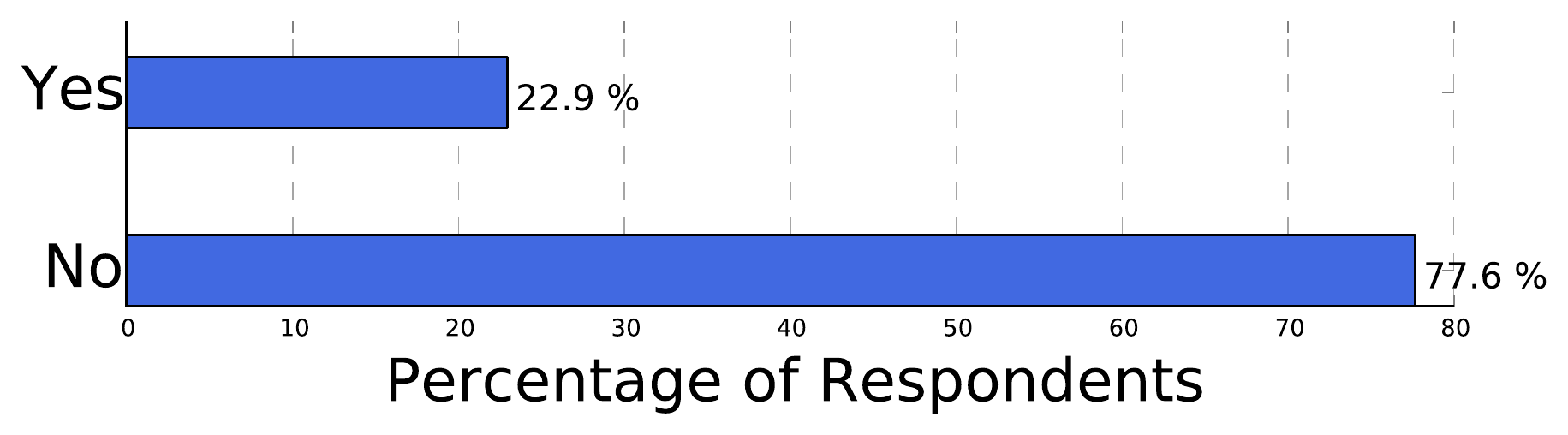}
  \caption{Are there any instances in which the center stack distracts you?}
  \label{fig:10}
  \end{center}
\end{figure}

This appendix describes the full results of the instrument cluster questionnaire. Fig.~\ref{fig:1} shows the results of the question "What information do you use while you drive?" where participants could choose multiple answers. We provided options, but added an "other" answer where participants had to specify if chosen. Fig.~\ref{fig:5} shows the results of the question "How do you ensure you are not speeding while driving?". The majority of respondents responded with "Looking at the instrument cluster". Furthermore, Fig.~\ref{fig:7} most participants don't think it is helpful to have all instrument cluster related information in the center console. Fig.~\ref{fig:8} and Fig.~\ref{fig:8h} describe the responses to the question "How much would you pay to upgrade the Tesla Model 3 with a traditionally placed instrument cluster or HUD?". The result of the last two questions about driver distraction are shown in Fig.~\ref{fig:9} and Fig.~\ref{fig:10}.

\section{Appendix B}\label{sec:appendix}
This appendix shows the absolute and mean estimation error of all participants and their average for the second speed estimation experiment.

\newcommand{\errorwidth}{0.95\textwidth}

\begin{figure*}
\centering
   \begin{subfigure}[b]{\errorwidth}
   \includegraphics[width=1\linewidth]{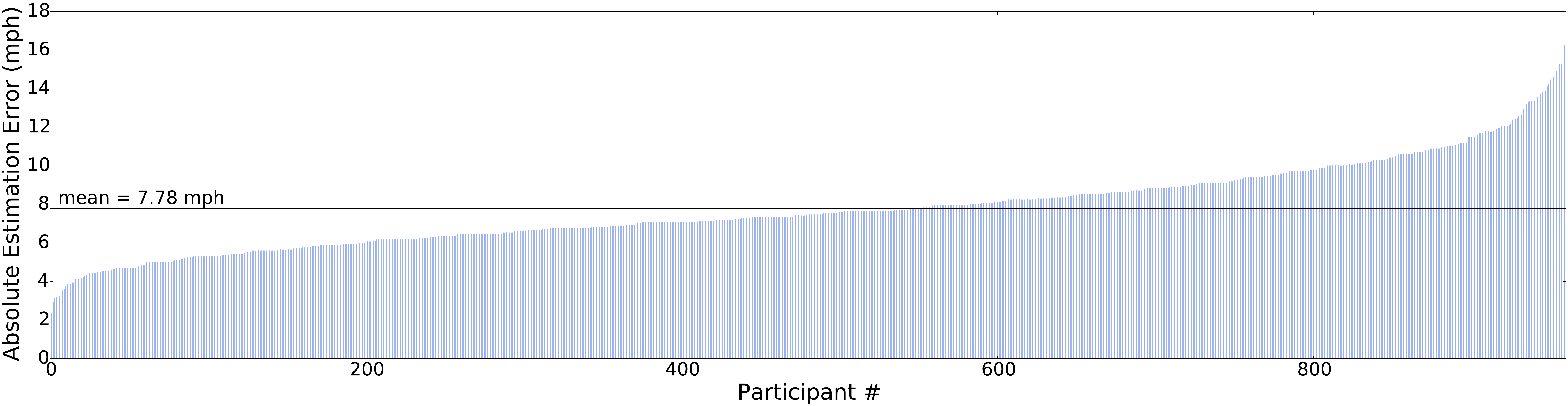}
   \caption{}
   \label{fig:abs} 
\end{subfigure}

\begin{subfigure}[b]{\errorwidth}
   \includegraphics[width=1\linewidth]{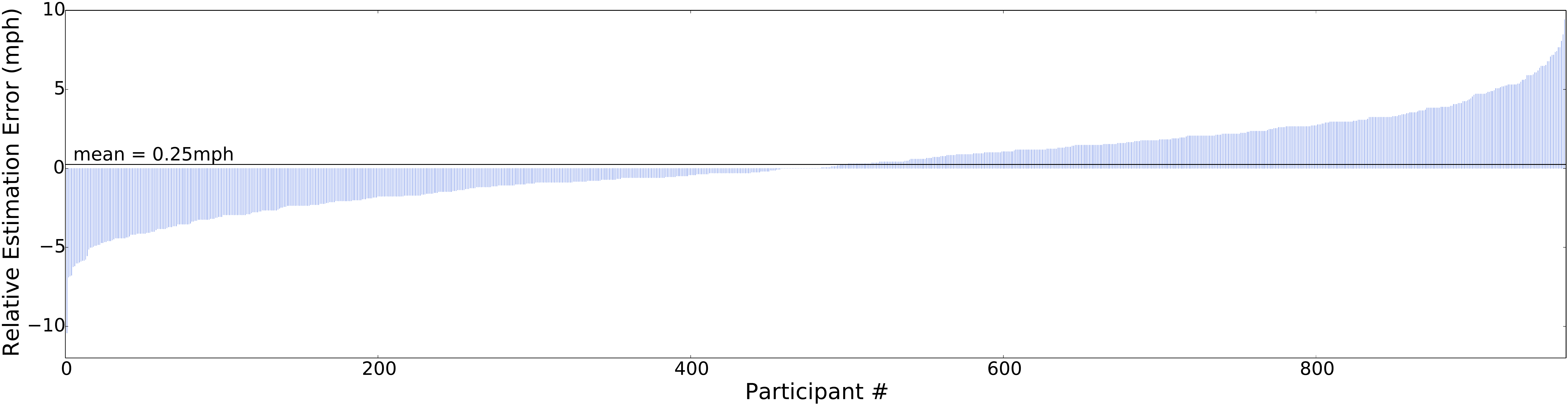}
   \caption{}
   \label{fig:rel}
\end{subfigure}

\caption[]{This chart shows the absolute estimation error per participant (a) and the relative estimation error for each participant for the second speed estimation experiment. The line indicates the mean estimation error of all participants.}
\end{figure*}

\begin{figure*}
\centering
  \begin{subfigure}[b]{\errorwidth}
   \includegraphics[width=1\linewidth]{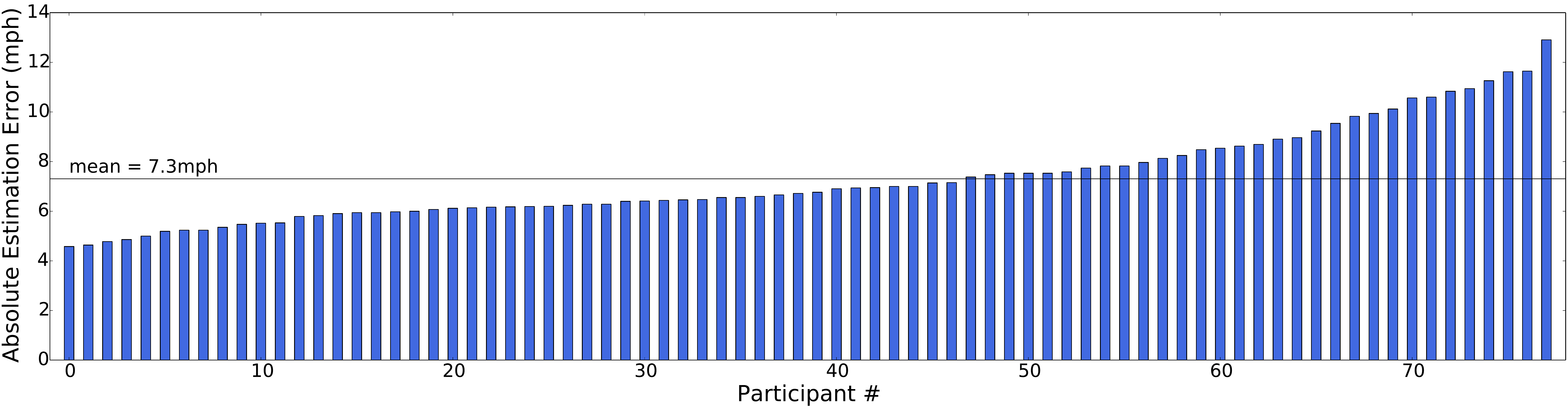}
   \caption{}
   \label{fig:abs} 
\end{subfigure}

\begin{subfigure}[b]{\errorwidth}
   \includegraphics[width=1\linewidth]{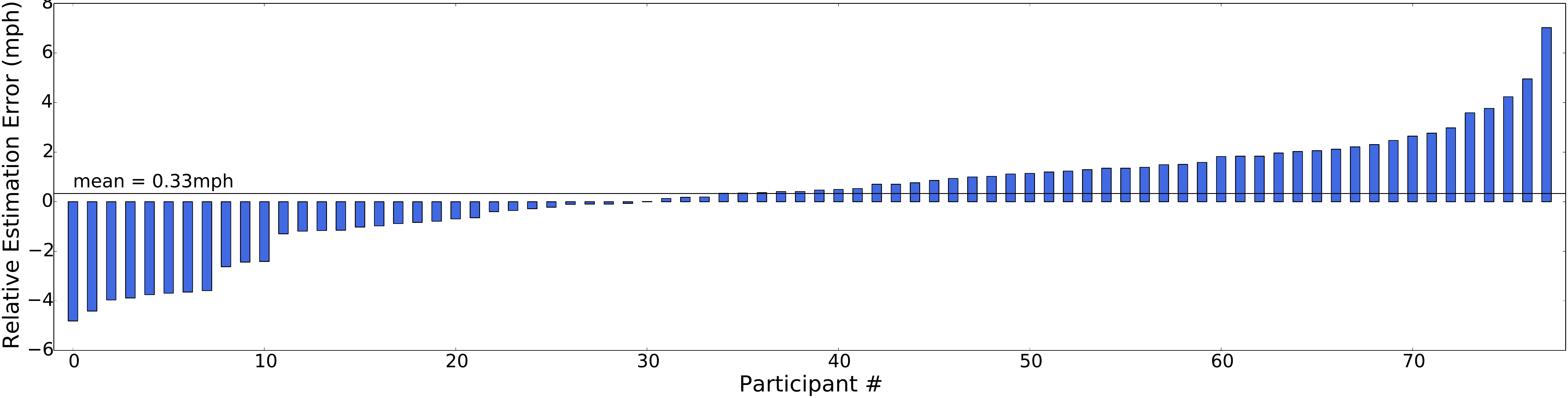}
   \caption{}
   \label{fig:rel}
\end{subfigure}

\caption[]{This chart shows the absolute estimation error per participant (a) and the relative estimation error for each participant for the second speed estimation experiment. The line indicates the mean estimation error of all participants.}
\end{figure*}

\end{document}